# CVD graphene-PMMA nanolaminates for flexible gas barrier

Antonio Baldanza[1], Maria Giovanna Pastore Carbone[2], Cosimo Brondi[1], Anastasios C. Manikas[3], Giuseppe Mensitieri[1,*], Christos Pavlou[2], Giuseppe Scherillo[1], Costas Galiotis[2,3,*]

[1] Department of Chemical, Materials and Production engineering, University of Naples Federico II, P.leTecchio 80, 80125 Naples, Italy; antonio.baldanza@unina.it, cosimo.brondi@unina.it, giuseppe.mensitieri@unina.it, giuseppe.scherillo@unina.it
[2] Institute of Chemical Engineering Sciences, Foundation for Research and Technology – Hellas (FORTH/ ICE-HT), 26504 Patras, Greece
[3] Department of Chemical Engineering, University of Patras, 26504 Patras, Greece; e-mail@e-mail.com
\* Correspondence:
mensitie@unina.it; G.M.
galiotis@chemeng.upatras.gr; C.G.

**Abstract:** Successful ways to fully exploit the excellent structural and multifunctional performance of graphene and related materials are of great scientific and technological interest. New opportunities are provided by the fabrication of a novel class of nanocomposites, with a nanolaminate architecture. In this work, by using the iterative lift-off/ float-on process combined with wet depositions, we have incorporated cm-size graphene monolayers produced via Chemical Vapour Deposition into a PMMA matrix with a controlled, alternant-layered structure. The produced nanolaminate shows a significant improvement of mechanical properties, with enhanced stiffness, strength and toughness, with the addition of only 0.06 vol% of graphene. Furthermore, oxygen and carbon dioxide permeability measurements performed at different R.H. levels, reveal that the addition of graphene leads to significant reduction of permeability, compared to neat PMMA. Overall, we demonstrate that the produced graphene-PMMA nanolaminate surpasses, in terms of gas barrier properties, the traditional discontinuous graphene-particle composites, with a similar filler content. Moreover, we have highlighted an interesting non-monotonic behavior of the gas permeability of a single graphene layer as a function of relative humidity. This works suggests possible use of CVD graphene-polymer nanolaminates as flexible gas barrier thus enlarging the spectrum of applications for this novel material.

## 1. Introduction

Two-dimensional (2D) crystals are the most promising materials to date, thanks to their extraordinary physicochemical properties and their potential to replace conventional ones in several applications [1]. Graphene, for instance, exhibits outstanding mechanical, electronic, optical and barrier properties, and is a promising candidate for many applications, from flexible electronics to composites and coatings [1,2]. The characteristic two-dimensional shape of graphene and related materials triggers these unique properties such as, among others, the capability to block small penetrant molecules [3]. Excellent results have been recently reported in the literature regarding the mass transport of low molecular weight compounds (penetrants) through graphene and other 2D materials [4]. It has been found that pristine, exfoliated graphene is impermeable to several penetrants, also including helium [5,6]. In fact, due to the high electronic density of the carbon rings, graphene can repulse atoms, gas molecules as well as water, leading to an intrinsic extremely low permeability of the defect-free layer [7]. However, macroscale sheets of perfect graphene without defects are impossible to obtain by mechanical exfoliation as this technique only yields micron-sized flakes [1].

Most of the attempts reported in the literature towards practical application of graphene as gas barrier concern the fabrication of nanocomposite membranes or coatings, which are obtained by dispersing liquid or chemically exfoliated graphene flakes and related materials (GRMs) – such as graphene nanoplatelets (GNPs), graphene oxide (GO) and reduced graphene oxide (rGO) – into a polymer matrix [7-9]. In principle, a homogeneous dispersion of impermeable graphene nanosheets results in a depleted gas permeability of the graphene-polymer nanocomposites due to the increased diffusion path of the permeant [10]. Also, the barrier properties of the composite depend on the aspect ratio (AR), the loading, the degree of dispersion and the orientation of the graphene nanosheet as well as on the structure organization of the polymeric chains at the interface with the graphene nanosheets [11]. Overall, the incorporation of GRMs in the form of discontinuous nanosheets into polymer matrices has been found to contribute to significant enhancement of gas barrier properties [7]. However, it is important to note that low loadings of graphene nanosheets can even increase the permeability of polymers to gases [12,13]; on the other hand, nanocomposites with higher GRM loadings that are effective for gas barrier applications may suffer from embrittlement due to poor dispersion of the nanosheets and aggregation phenomena, thus creating structural defects or deteriorating the mechanical properties of the polymer [14].

In general, the typical structure of the discontinuous graphene-polymer composites lacks the precise architecture and 2D nano-porous structure, which is characteristic of all-graphene, highly packed membranes [8]. In principle, the optimization of barrier properties can be achieved (with the lowest amount of nanomaterial) by producing films of alternated graphene and polymer layers [15]. In such multilayer membranes, a crucial role is played by the distribution of defects (nanometric holes and/or microcracks) onto the graphene layer which provide the prevalent permeation paths to the penetrant molecules in passing through [16,17]. To this regard, it is important to remark that it is not feasible to produce GRM-based nanocomposites as well as multilayer in which the nanosheets are free of defects [18,19]. In fact, according to the statistical thermodynamics, mono- and multi-vacancies on the 2D graphene lattice are intrinsically present for entropic reasons at T > 0 K [19]. Moreover, additional defects of single graphene layer can be induced by the manufacturing process [18,19]. The distribution of size and shape of these voids directly governs the permeability mechanism of the graphene nanosheets and, in turn, the barrier properties performances as well as the gas permeation selectivity of the nanocomposites [9].

The production of high quality, continuous, macroscopic 2D materials with controlled layer thickness and reduced number of defects is a growing field, which has garnered considerable attention [15] and has recently provided new opportunities for the preparation of effective composites with unprecedented structural and multifunctional properties. Chemical Vapour Deposition (CVD), for instance, has been widely utilized to synthesize wafer-scale 2DM and represents a good compromise between quality, quantity and cost. Recently, some of the authors have proposed a novel way to incorporate cm-size CVD graphene monolayer into a polymer matrix with a controlled, nanolaminate architecture. The obtained CVD graphene-poly (methyl methacrylate) (PMMA) nanolaminate membranes were found to present enhanced mechanical, electrical and EMI shielding properties, with record values in the state-of-the-art [20]. Herein, we demonstrate that the controlled incorporation of CVD graphene into the nanolaminate configuration not only offers improved gas barrier properties compared to the polymer matrix but outperforms traditional discontinuous GRM composites, for the same graphene content. To this regard, steady state permeation measurements have been performed with two typical gases of interest (namely, $CO_2$ and $O_2$) for gas barrier applications, at different levels of water relative humidity (R.H.).

## 2. Materials and Methods

*2.1 Material preparation*

Nanolaminate membranes were produced by iterative transfer of graphene-PMMA (Gr-PMMA) layers using the 'lift-off/ float-on' process combined with wet depositions described in [20]. Single-layer graphene was grown on copper foil (JX Nippon Mining & Metals, 35 μm-thick, 99.95%) in a commercially available CVD reactor (AIXTRON Black Magic Pro, Germany). A commercial solution of polymethyl methacrylate (PMMA) (495 PMMA A6, Microchem Corp.) was adopted for the fabrication of the polymer matrix. Each Gr-/PMMA layer was fabricated by spin coating the PMMA solution on the CVD graphene on Cu foil, at 2000 RPM for 1 min at RT. The copper was then etched away with an aqueous solution of ammonium persulphate (APS) and the remaining Gr-PMMA layer was rinsed with distilled water. By slowly reducing the water level, the floating Gr-PMMA layer was then deposited on another Gr-PMMA layer on a Cu foil, which represents the substrate for subsequent depositions. After each deposition, the multi-layer was firstly dried at 40 °C under vacuum to remove the excess of water and then post-baked at 150 °C for a few minutes, to improve adhesion between subsequent layers. The cycle was repeated until the desired number of Gr-PMMA layers was reached (20). At the end, the Cu substrate was etched away in APS solution to release the freestanding Gr-PMMA nanolaminate, then rinsed with water and finally dried at 40 °C under vacuum to remove the excess of water. A similar procedure was adopted to produce a control nanolaminate of neat PMMA, in which bare Cu foils (without graphene) were used as sacrificial substrates.

*2.2 Thickness measurements for the evaluation of graphene volume fraction*

The thickness of the single Gr-PMMA layer ($t_{Gr-PMMA}$) was measured using the scratch test method, through atomic force microscopy (AFM) [21]. After etching the Cu foil, the layer was deposited on Si wafer and was scratched using a scalpel without damaging the substrate. AFM images of the scratch were acquired using a Dimension Icon (Bruker), in the Peak Force Tapping mode using ScanAsyst-Air probes (stiffness 0.2−0.8 N/m, frequency ~80 kHz). The average depth of the scratch below the mean surface plane, corresponding to the film thickness, was evaluated using the cross-section analysis of the Nanoscope Analysis software. Several scratches were measured for each deposited layer to allow statistical analysis of data.

The final thickness of the produced Gr-PMMA nanolaminate was determined as the mean of 10 measurements with a digital micro-meter with a resolution of 0.1 μm (Mitutoyo, Japan).

*2.3 Raman spectroscopy*

Raman spectroscopic mapping has been performed on the nanolaminate with a Renishaw Invia Raman Spectrometer on an area of 100 × 100 μm², by acquiring spectra at steps of 3 μm, using a 514 nm laser excitation and a 100× lens. Laser power on the specimens was kept below 1 mW to avoid laser-induced heating. Raman spectra were baseline corrected and characteristic peaks of graphene were fitted to Lorentzian functions, the spectroscopic parameters (peak position and full width at half maximum, FWHM) being recorded at each position of the mapping.

*2.4 Uniaxial tensile testing*

Tensile tests on the nanolaminates were performed using a micro-tensile tester (MT-200, Deben UK Ltd) equipped with a 5 N load cell. Rectangular specimens with a gauge length of 25 mm and a width of 1 mm were secured onto paper testing cards using a two-part cold curing epoxy resin. The specimens were then loaded in tension with a crosshead displacement speed of 0.2 mm min$^{-1}$. Stress and strain were calculated based on the measured machine-recorded forces and displacements and, for each specimen, the Young's modulus was estimated through a linear regression analysis of the initial linear portion of the stress–strain curves (up to ~0.4% strain). The estimation of both the Young's modulus and tensile strength was obtained by the mean values of at least ten specimens, and the experimental errors are the deviation from the mean values.

*2.5 Scanning Electron Microscopy*

Scanning Electron Microscopy (ZEISS SUPRA 35VP SEM) was employed to assess the morphology of the produced nanolaminates.

*2.6 Permeability measurements*

Permeabilities of oxygen and carbon dioxide through nanolaminate films were measured at different R.H. levels using a commercial permeabilimeter (Multiperm by ExtraSolutions S.R.L. Pisa, Italy). Both the sides of the nanolaminate are exposed to gas streams at a total pressure of 1 atm so that any convective contribution to the gas transport through the membrane can be disregarded and the whole measurement chamber is maintained at a controlled temperature (25°C in the case at hand). To remove any low molecular weight compounds possibly absorbed within the membrane, before each permeation test both sides of the film are flushed with pure nitrogen (service gas) for a time span long at least 5 times the expected transient of the permeation experiment. Each permeation test starts by switching the gas stream on the upstream side from the service gas to the testing gas ($CO_2$ or $O_2$ in the case at hand) at a fixed R.H. Simultaneously, the same R.H. condition is also imposed on the downstream side for the service gas (nitrogen) flow, thus realizing symmetric R.H. conditions. Once the experiment is started, the service gas on the downstream side continuously removes the permeated test gas molecules. The concentration of test gas molecules within the service gas stream is measured in situ by using an electrochemical sensor in the case of oxygen and an infrared sensor in the case of carbon dioxide. Once steady state permeation conditions are attained (i.e., constant permeation flux through the membrane and, in turn, constant concentration of test gas within the service gas stream on the downstream side) the steady state flux of the test gas through the membrane, $J_{ss}$, is estimated. Then, from this value, the permeability coefficient, P, of the test gas is calculated according to the following equation:

$$P = \frac{J_{SS}}{\left(\Delta p / t\right)} \qquad (1)$$

where $l$ represents the membrane thickness and $\Delta p$ stands for the difference of the partial pressure of the test gas between the upstream and the downstream sides. The partial pressure of the test gas on the downstream side is very close to 0. All these calculations are automatically performed by the software governing the apparatus. In the operating conditions of the experiments, since the nitrogen solubility is much smaller than those of oxygen, carbon dioxide and water [22-25], effects of nitrogen on the measured permeabilities can be neglected.

## 3. Results and Discussion

*3.1 Characterization of materials*

By using the all-fluidic 'lift-off/ float-on' process schematically depicted in Figure 1a, a freestanding nanolaminate consisting of twenty graphene-PMMA (Gr-PMMA) layers was produced. In the nanolaminate configuration, the volume fraction of graphene ($V_{Gr}$) is defined as:

$$V_{Gr} = \frac{t_{Gr}}{t_{Gr} + t_P} \quad (2)$$

being $t_{Gr}$ the thickness of monolayer graphene (0.334 nm) and $t_P$ the thickness of the polymeric layer. Therefore, to estimate the content of graphene in the produced nanolaminate, a preliminary investigation of the thickness of the single Gr-PMMA layer has been performed. According to the scratch test method, the thickness of the single Gr-PMMA layer ($t_{Gr-PMMA}$) is ca. 550 nm (Figure 1b), indicating that the $t_{PMMA} \cong 549.5$ nm and that the nominal volume fraction of graphene in the nanolaminate was 0.06%. This value well matches the actual volume fraction as inferred inversely from the final thickness of the nanolaminate which has been measured by a digital micro-meter. A representative SEM image of the cross-section of the specimen is shown in Figure 1c and clearly reveals the stratified architecture of the produced membrane, highlighting a regular lamination sequence of the Gr-PMMA layers.

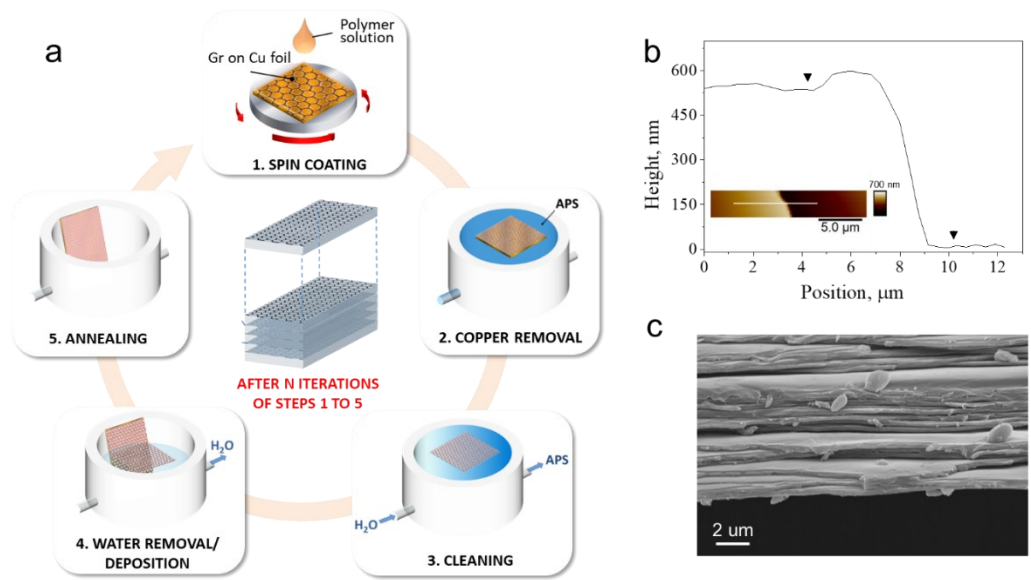

**Figure 1.** (a) Schematic illustration of the iterative 'lift-off/ float-on' process combined with wet depositions adopted to produce the Gr-PMMA nanolaminates. (b) Thickness evaluation of the single Gr-PMMA layer deposited on a Si wafer: representative cross-section of the scratch and AFM image as inset. (c) SEM image in the cross-section plane of the nanolaminate.

Raman spectroscopy was employed to assess the quality of the graphene layers incorporated in the nanolaminate. The representative Raman spectrum acquired from the produced nanolaminate is shown in Figure 2a and presents the typical spectroscopic features of graphene, namely the graphitic (G) and the strong second-order (2D) peaks [26]. The homogeneous distribution of these features within the specimen is revealed in the contour plots shown in Figure 2b, thus confirming the successful incorporation of continuous graphene in the nanolaminate. In particular, the G peak located at ~1592 cm$^{-1}$ and the 2D peak located at ~2706 cm$^{-1}$ reveal that graphene experiences a slight residual compression. The Int(2D)/Int(G) ratio larger than 2 indicates the monolayer feature of graphene which is retained after multiple depositions; however, the non-zero but small Int(D)/Int(G) indicates the presence of minor structural defects that are likely induced during the manipulation of the layers in the lamination process (Figure 2c).

The mechanical performance of the produced nanolaminate has been assessed by means of uniaxial tensile testing and representative stress strain curves for the Gr-PMMA nanolaminate, as well as for the PMMA control sample, are shown in Figure 3. It is interesting to note that the addition of graphene in the nanolaminate results in a substantial increase of the Young's modulus, from 1.8 ± 0.2 MPa to 2.4 ± 0.2 MPa. Also, the tensile strength is found to increase by 100% ca., compared to the control sample. By following the approach adopted by Vlassiouk et al. [27], the use of a simple rule of mixture allows to estimate the effective contribution of graphene to the modulus and the strength of the nanolaminate, which yield 1 TPa and 40 GPa, respectively. This unprecedented improvement of the mechanical properties in tension has been already ascribed to the efficient reinforcement provided by the large-size graphene sheets in the nanolaminate architecture [20]. Furthermore, it is also worth noting that the nanolaminate shows a tougher behaviour, with two-fold increase of elongation at break, compared to the neat PMMA. This can be likely ascribed to possible phenomena of slippage of the graphene layers upon large deformations, as already observed by Liu et al [28].

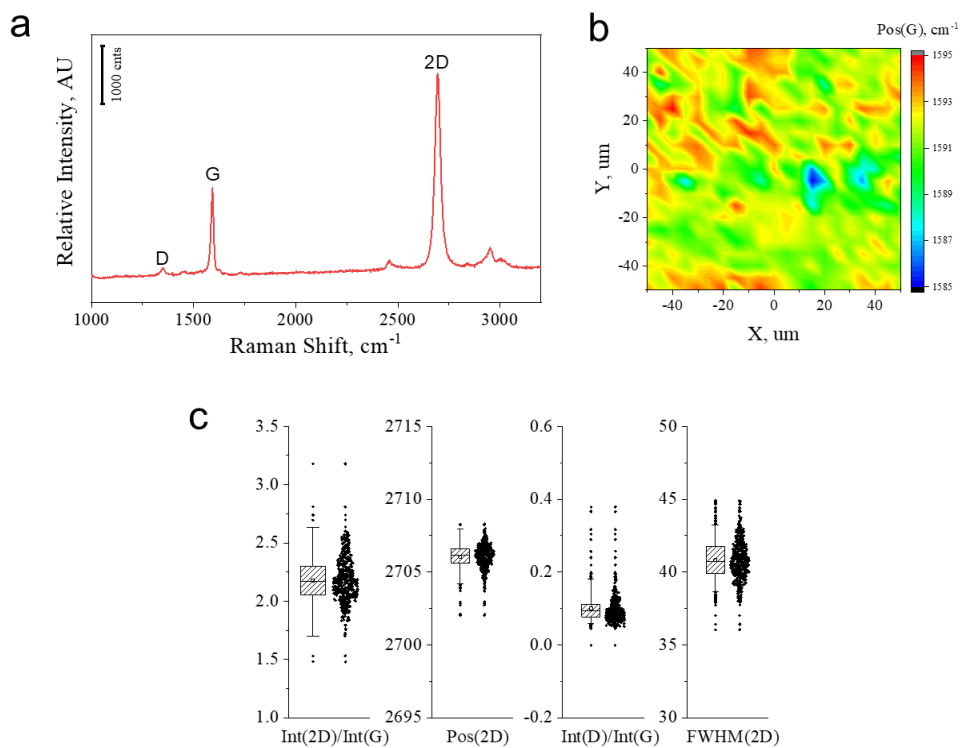

**Figure 2.** (a) Representative Raman spectra collected from the Gr-PMMA nanolaminate under investigation. (b) Contour map of the position of the G peak. (c) Box plots for Int(2D)/Int(G), Pos(2D), Int(D)/Int(G) and.FWHM(2D).

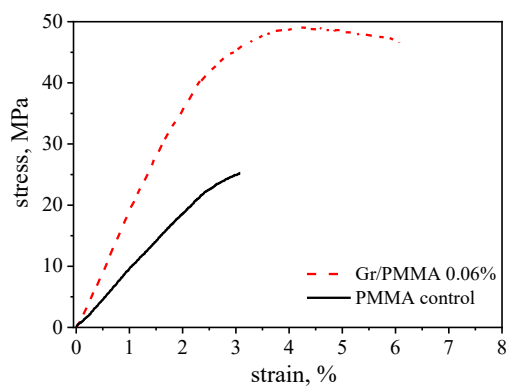

**Figure 3.** Representative stress-strain curve of the Gr-PMMA nanolaminate (red dotted line) and of the control PMMA sample (black solid line) obtained by uniaxial tensile testing.

*3.2 Results of permeability measurements*

Permeability tests of gaseous $CO_2$ and $O_2$ through the nanolaminate films were performed at 25°C at 1 atm. Experiments were conducted at three RH values (0, 50 and 80%), symmetrically imposed at upstream and downstream sides of the films.

In Figure 4a,b are reported bar diagrams of the $CO_2$ and $O_2$ (pure and humidified) permeability coefficients for the pure PMMA nanolaminate and the Gr-PMMA nanolaminate membranes. The corresponding numerical values are also reported in Tables 1 and 2, respectively. We remark that the permeability coefficients of the pure gases in neat PMMA nanolaminate agree with the values reported in literature for commercial amorphous PMMA [22,23,29,30]. This indicates that the lamination process, per se, does not induce relevant effects on the bulk structure of PMMA layers within the nanolaminate. The enhanced $CO_2$ and $O_2$ barrier properties measured in the whole range of investigated R.H.s and discussed below are to be mainly ascribed to the addition of graphene layers since the morphological investigations performed on the nanocomposite showed that the bulk of PMMA layers do not exhibit evident orientation effects as well as crystallization effects induced by the contact of macromolecules with graphene layers that may affect the value of the permeability coefficient [31].

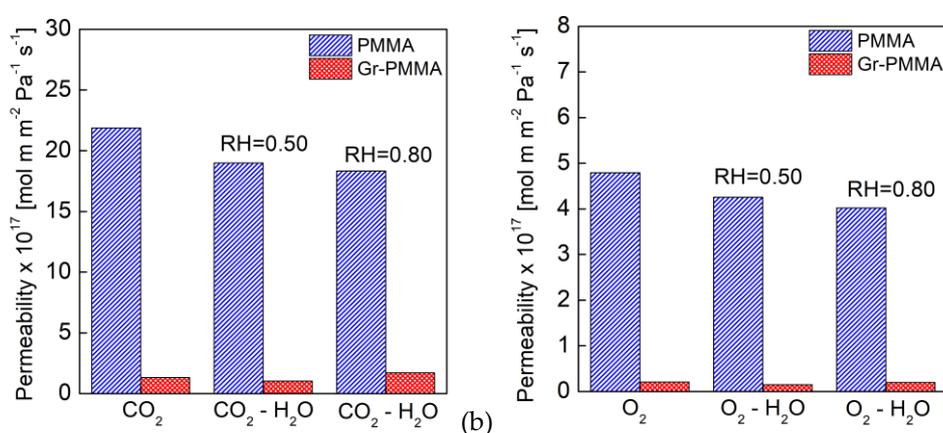

**Figure 4**. Gas permeability coefficients at 25°C through PMMA (blue bars) and Gr-PMMA (red bars) for (a) $CO_2$ and humidified $CO_2$ and for (b) $O_2$ and humidified $O_2$.

**Table 1.** Permeability coefficients of $CO_2$ through the nanolaminates at different R.H. levels.

| Nanolaminate/*Permeating Gas* | $P \cdot 10^{17}$ [mol·m·m$^{-2}$·Pa$^{-1}$·s$^{-1}$] |
|---|---|
| PMMA/*CO$_2$* | 21.88 |
| Gr-PMMA/*CO$_2$* | 1.32 |
| PMMA/*CO$_2$-H$_2$O @RH=0.50* | 18.99 |
| Gr-PMMA/*CO$_2$-H$_2$O @RH=0.50* | 1.03 |
| PMMA/*CO$_2$-H$_2$O @RH=0.80* | 18.32 |
| Gr-PMMA/*CO$_2$-H$_2$O @RH=0.80* | 1.71 |

**Table 2.** Permeability coefficients of $O_2$ through the nanolaminates at different R.H. levels.

| Nanolaminate/*Permeating Gas* | $P \cdot 10^{17}$ [mol·m·m$^{-2}$·Pa$^{-1}$·s$^{-1}$] |
|---|---|
| PMMA/*O$_2$* | 4.79 |
| Gr-PMMA/*O$_2$* | 0.21 |
| PMMA/*O$_2$-H$_2$O @RH=0.50* | 4.25 |
| Gr-PMMA/*O$_2$-H$_2$O @RH=0.50* | 0.15 |
| PMMA/*O$_2$-H$_2$O @RH=0.80* | 4.02 |
| Gr-PMMA/*O$_2$-H$_2$O @RH=0.80* | 0.20 |

Several literature contributions report an improvement of barrier properties of PMMA after the incorporation of GRM in the form of flakes/nanosheets [7]. For instance, Morimune et al. have observed a significant improvement of barrier properties in discontinuous PMMA/GO nanocomposites following the addition of 1 wt% (0.5 vol%) of µm-size GO to the PMMA matrix leading to a decrease of the $O_2$ permeability by 50% [32]. Chang et al. prepared PMMA nanocomposites filled with carboxyl–graphene nanosheets and found that, compared to neat PMMA, the nanocomposite membranes with 0.5 wt% (0.25 vol%) carboxyl–graphene nanosheets loading show about 70% and 61% reduction in $O_2$ and $H_2O$ permeability, respectively [33].

Indeed, the nanolaminate investigated here shows a higher reduction of the $O_2$ permeability as compared to the neat PMMA (ca 95%), which is achieved with the addition of only 0.06 vol% of graphene. Such enhancement of the barrier properties is to be ascribed to the controlled and homogeneous dispersion of the graphene layers with high aspect ratio within the PMMA matrix, which is achieved within the alternated-layered structure of the nanolaminate. In fact, it has been theoretically estimated and experimentally demonstrated that, when graphene sheets possess an aspect ratio higher than 1000, a tiny volume fraction of graphene allows to achieve a significantly improved barrier performance ($P_{nanocomposite}/P_{polymer} \cong 0.1$) [31]. Based on the Nielsen approximation [10] for polymer nanocomposites with layered 2D sheets oriented perpendicularly to the diffusion direction, when the aspect ratio approaches infinity (as for the cm-size, atomic-thick graphene used in the nanolaminate at hand), the required amount of the filler materials to achieve a good barrier property is vanishingly small.

It is important to underline, however, that large-scale graphene monolayers produced via CVD contain numerous defects (e.g., vacancies, grain boundaries, wrinkles) that are originated during the synthesis and the transfer/deposition processes, and therefore its barrier properties are lower as compared to pristine graphene [34].

In view of these observations, a reasonable physical picture for the permeation process of the penetrant molecules through the polymer-graphene-based multilayer nanolaminate systems consists in assuming a series mechanism involving the ordinary diffusive behavior within the polymer-penetrant phase and diffusive jumps through graphene defects (voids, cracks) statistically distributed onto each graphene monolayer within the composite [8,9,35]. According to this framework, the permeability coefficient of each penetrant through the nanolaminate membrane, $P_{Gr-PMMA}$, can be expressed by the following equation [8,9,36]:

$$\frac{t_{Gr-PMMA}}{P_{Gr-PMMA}} = \sum_{i=1}^{N} t_{Gr}/P_{i,Gr} + \sum_{j=1}^{M} t_{PMMA}/P_{j,PMMA} = \frac{Nt_{Gr}}{P_{Gr}} + \frac{Mt_{PMMA}}{P_{PMMA}} \quad (3)$$

where $t_{Gr}$ and $t_{PMMA}$ represent respectively the single graphene layer and the single PMMA layer thickness and $N$ and $M$ stand for the total number of graphene and PMMA layers forming the nanolaminate, respectively (in our case, $N=M=20$). Finally, $P_{i,Gr}$ and $P_{j,PMMA}$ stand for the testing gas permeability coefficient through respectively the $i$-th and $j$-th graphene and PMMA layer. Due to the low values of penetrant concentration which is expected to occur within the polymeric phases [25], one can reasonably assume that $P_{j,PMMA}$ does not depend significantly upon the position of the specific $j$-th layer. Moreover, being the number of PMMA/Gr layer sandwiches sufficiently high to allow a reliable statistical "averaged" approach, one can also assume a unique "average" value for $P_{i,Gr}$ within the system, so that the second equality in eq. (3) holds true. To better elucidate the role of the distribution of the defects size present into the graphene layers on the permeation mechanism, one can start to analyze the case of the two pure gases investigated. To this regard, eq. (3) can be used to estimate the "average" permeability, $P_{Gr}$, of each pure gas through the graphene layers within the nanocomposite. In fact, given the number of graphene and of PMMA layers, the gas permeability through the Gr-PMMA nanocomposite and through the pure PMMA nanolaminates (see Tables 1 and 2), $P_{Gr}$ can be easily calculated from the following expression:

$$P_{Gr} = \frac{N \cdot t_{Gr}}{\frac{t_{Gr-PMMA}}{P_{Gr-PMMA}} - \frac{M \cdot t_{PMMA}}{P_{PMMA}}} \quad (4)$$

One obtains a value of $P_{Gr}$ equal to $8.53 \cdot 10^{-21}$ [mol m m$^{-2}$ Pa$^{-1}$ s$^{-1}$] in the case of $CO_2$ and to $1.33 \cdot 10^{-21}$ [mol m m$^{-2}$ Pa$^{-1}$ s$^{-1}$] in the case of $O_2$. The ratio between the pure $CO_2$ permeability of a graphene layer and that of pure $O_2$ is higher than its value for pure PMMA (6,40 vs. 4,57), thus suggesting that the presence of the graphene layers introduces a molecular 'sieving' mechanism in the nanocomposite ruled by the graphene layer defects size, that is not present in the glassy PMMA that is characterized by a solubility selectivity more than by a size selectivity [35] in view of its high free volume. To support this hypothesis, one can observe that the values of the kinetic diameters, taken from the literature [37], are respectively equal to 0.330 nm for $CO_2$ and 0.346 nm for $O_2$ which are in line with the higher $P_{Gr}$ observed for $CO_2$.

In general, the gas permeability coefficient of a multilayer membrane, $P$, can be properly quantified according to the following phenomenological equation, which has been already adopted by Pierleoni et. al. [9] in dealing with pure gas permeabilities through different GO-polymeric multilayer nanocomposite membranes:

$$P = a \cdot e^{-b \cdot k_d} \tag{5}$$

where $k_d$ represents the kinetic diameter [9,37], and the parameter $b$ is simply retrieved from the slope of the corresponding permeability coefficients curve (semilog scale) versus the corresponding molecular kinetic diameter and provides a quantitative estimation of the "membrane sieving capacity".

In this context, it is of interest to compare the values of $b$ for pure PMMA and for Gr-PMMA nanolaminates in the case of $CO_2$ and $O_2$ with the "average" value of $b$ for a single graphene layer. The corresponding values are $b$ = 94.94 nm$^{-1}$ in the case of pure PMMA, $b$ = 114.89 nm$^{-1}$ in the case of the PMMA-graphene nanolaminate and $b$ = 115.98 nm$^{-1}$ for the single graphene layer within the nanocomposite. Indeed, the value of $b$ in the case of the nanocomposite is higher than that of the nanolaminate made of neat PMMA and slightly lower than the 'average' $b$ of the single graphene layer. This result is consistent with the described physical picture, pointing out that the 'necking factor' for the penetrants permeation mechanism is given by jumps through voids of proper size and shape distributed onto each graphene monolayer.

The stronger dependence of the permeability on the penetrant molecular size, compared to the pure PMMA nanolaminate membrane, which is observed in the case of the single graphene monolayer, and hence for the nanocomposite, suggests that most of the graphene layers would mainly display nano-sized defects.

The role of graphene defects on the penetrant permeation mechanism in the nanocomposite system is expected to be significant also in the case of gas permeation in the presence of water vapor. To elucidate possible effects of water molecules on the mass transport mechanism of $CO_2$ and $O_2$, we have investigated the permeation of the two gases at different R.H. levels.

We first analyzed the case of the pure PMMA nanolaminate film, finding that both the $CO_2$ and $O_2$ permeability coefficients decrease as a function of R.H. level. Starting from the analysis of pure gases, the steady-state permeation coefficient $P$ can be expressed, in principle, as $P = \overline{D} \times \overline{S}$, where $\overline{D}$ and $\overline{S}$ represent, respectively an average thermodynamic diffusion coefficient and an average solubility coefficient [36,38]. In fact, both diffusivity and solubility coefficients can be a function of-penetrant concentration [38] and hence of the spatial position along the film thickness. In the case at hand, in the range of pressure considered, both diffusivity and solubility coefficients for $CO_2$ and $O_2$ can be assumed to be constant.

In the case of humidified streams, water vapor is absorbed within the PMMA membrane thus affecting both the diffusivity and solubility of the test gas. In particular, a monotonic decrease of permeabilities of both gases with RH is observed. The prominent effect is expected to be the reduction of solubility of the gases in view of the higher condensability of water. In fact, water competes with permanent gases for adsorption on excess free volume microvoids characteristic of glassy polymer [39]. In contrast, in the case of Gr-PMMA nanolaminate a non-monotonic effect of RH on test gas permeabilities has been detected. In fact, experimentally determined permeabilities of both $CO_2$ and $O_2$ display a minimum value at RH=0.5. To isolate the contribution of a single Graphene layer, we estimated its apparent permeabilities by using eq. (4) as applied to the case of RH=0.5 and RH = 0.8. The results are reported in table 3. Quite interestingly, both $CO_2$ and $O_2$ permeabilities for a single graphene layer also display such a non-monotonic behavior. The presence of water affects the permeabilities of gases through a graphene layer and the permeabilities first decrease, at a relative humidity value of 0.5, and then increase again at a relative humidity value of 0.8. Although lack of detailed information on the structure of graphene layers does not allow for an explanation of such an effect, this interesting result is likely related to the interplay between water blockage of passageways through the graphene layer and alteration of the layer structure induced by the presence of water molecules.

**Table 3.** Estimated permeability coefficients of $CO_2$ and $O_2$ through a graphene nanolayer at different R.H. levels.

| Permeating Gas | $P \cdot 10^{21}$ [mol·m·m$^{-2}$·Pa$^{-1}$·s$^{-1}$] |
|---|---|
| $CO_2$ | 8.53 |
| $CO_2$ @ RH=0.5 | 6.62 |
| $CO_2$ @ RH=0.8 | 11 |
| $O_2$ | 1.33 |
| $O_2$ @ RH=0.5 | 0.95 |
| $O_2$ @ RH=0.8 | 1.28 |

## 4. Conclusions

Films with a nanolaminate architecture, made of controlled alternating PMMA layers and cm-size graphene monolayers were produced via Chemical Vapour Deposition and characterized for their mechanical and gas barrier properties. A significant improvement of mechanical properties as compared to the pure PMMA has been registered, with enhanced stiffness, strength and toughness, with the addition of only 0.06 vol% of graphene.

Oxygen and carbon dioxide permeability measurements, conducted at different R.H. levels, revealed that the addition of graphene monolayers leads to a significant improvement of the gas barrier properties as compared to neat PMMA, outperforming the traditional discontinuous graphene-particle composites, with a similar filler content. A size-sieving effect of the graphene monolayer has been evidenced and an interesting effect of the relative humidity on permeation properties has been highlighted. In fact, a non-monotonic behavior of the gas permeability of a single graphene layer as a function of relative humidity has been detected that has been tentatively attributed to the interplay between water blockage of passageways through the graphene layer and alteration of the layer structure induced by the presence of water molecules.

**Author Contributions:** Conceptualization, C.G., M-G.P.C. and G.M.; methodology, A.C.M., C.P.; formal analysis, A.B., C.B., G.S.; investigation, A.B.; data curation, M.-G.P.C., A.B.; writing—original draft preparation, G.S., M.-G.P.C.; writing—review and editing, G.M.; supervision, C.G.; project administration, X.X.; funding acquisition, Y.Y. All authors have read and agreed to the published version of the manuscript.

**Funding:** The authors acknowledge the financial support of the research project Graphene Core 3, GA: 881603.

**Acknowledgments:** Dr. George Trakakis is acknowledged for providing CVD graphene samples.

**Conflicts of Interest:** The authors declare no conflict of interest..